\documentclass[twocolumn,5p,times]{elsarticle}

\usepackage{amssymb}

\usepackage[modulo]{lineno}
\usepackage[usenames]{color}

\usepackage{nicefrac}
\usepackage{multirow} 
\usepackage{hyperref} 
\usepackage{units}
\usepackage{fixltx2e}  
\usepackage{dblfloatfix}
\usepackage{amsmath}
\usepackage{bm} 

\journal{Astroparticle Physics}

\begin{document}

\begin{frontmatter}

\title{Estimation of the number of muons with muon counters}

\author[ITEDA]{A. D. Supanitsky \corref{cor1}}
\cortext[cor1]{Corresponding author: daniel.supanitsky@iteda.cnea.gov.ar}
\address[ITEDA]{Instituto de Tecnolog\'ias en Detecci\'on y Astropart\'iculas (CNEA, CONICET, 
UNSAM), Centro At\'omico Constituyentes, San Mart\'in, 1650 Buenos Aires, Argentina.}

\begin{abstract}

The origin and nature of the cosmic rays is still uncertain. However, a big progress 
has been achieved in recent years due to the good quality data provided by current
and recent cosmic-rays observatories. The cosmic ray flux decreases very fast with 
energy in such a way that for energies $\gtrsim 10^{15}$ eV, the study of these very 
energetic particles is performed by using ground based detectors. These detectors are 
able to detect the atmospheric air showers generated by the cosmic rays as a consequence 
of their interactions with the molecules of the Earth's atmosphere. One of the most 
important observables that can help to understand the origin of the cosmic rays is 
the composition profile as a function of primary energy. Since the primary particle 
cannot be observed directly, its chemical composition has to be inferred from parameters 
of the showers that are very sensitive to the primary mass. The two parameters more 
sensitive to the composition of the primary are the atmospheric depth of the shower 
maximum and the muon content of the showers. Past and current cosmic-rays observatories 
have been using muon counters with the main purpose of measuring the muon content of 
the showers. Motivated by this fact, in this work we study in detail the estimation 
of the number of muons that hit a muon counter, which is limited by the number of 
segments of the counters and by the pile-up effect. We consider as study cases muon 
counters with segmentation corresponding to the underground muon detectors of the 
Pierre Auger Observatory that are currently taking data, and the one corresponding 
to the muon counters of the AGASA Observatory, which stopped taking data in 2004.        
 
\end{abstract}


\end{frontmatter}

\section{Introduction}

The cosmic ray energy spectrum extends over several orders of magnitude in 
energy. The highest primary energies observed at present are of the order 
of $10^{20}$ eV. Above $\sim 10^{15}$ eV the cosmic ray flux is so small 
that these very energetic particles are studied by means of ground-based 
detectors, which are able to detect the atmospheric air showers generated
by the cosmic ray interactions with the molecules of the atmosphere. Since
the primary particle is not observed directly, its energy, arrival direction,
and chemical composition have to be inferred from the shower information 
obtained by the detectors.   

The origin of the cosmic rays is still unknown. The three main observables 
used to study their nature are: The energy spectrum, the distribution
of their arrival directions, and the chemical composition. The composition of the 
primary particle has to be inferred from different properties of the showers. The
most sensitive parameters to the primary mass are the atmospheric depth at which 
the shower reaches its maximum development and the muon content of the showers 
\cite{Supa:08,Supa:09,Kampert:12}. In general, the muon density at a given 
distance to the shower axis is used in composition analyses.      

The composition profile as a function of energy is very important to understand 
several aspects of the cosmic-ray physics. In particular, at the highest energies
the composition information plays an important role to find the transition between 
the galactic and extragalactic components of the cosmic rays \cite{MTanco:07,Aloisio:12} 
and to elucidate the origin of the suppression observed at $\sim 10^{19.7}$ eV 
\cite{Kampert:13}. The composition analyses are subject to large systematic 
uncertainties originated by the lack of knowledge of the hadronic interactions at 
the highest energies (see for instance \cite{Supa:09a}). The composition is 
determined by comparing experimental data with simulations of the atmospheric showers
and the detectors (when it corresponds). The showers are simulated by using high-energy 
hadronic interaction models that extrapolate low-energy accelerator data to the highest
energies. This practice introduces large systematic uncertainties even when models 
updated using the Large Hadron Collider data are considered. Moreover, experimental 
evidence has been found recently about a muon deficit in shower simulations
\cite{Gesualdi:20,Cazon:19}. Even though this is an important limitation for 
composition analyses, it is expected that mass-sensitive parameters, obtained 
with the next generation of high-energy hadronic interaction models, present smaller 
differences allowing for a reduction of the systematic uncertainties introduced 
by those models. 
  
Past and current cosmic-rays experiments have been measuring muons by using 
different types of detectors \cite{Cazon:19}. A particular class of detector
is the muon counter. This type of detectors has been used in the past in 
the Akeno Giant Air Shower Array (AGASA) \cite{AGASA:95} and at present in 
the Pierre Auger Observatory \cite{AMIGA:19}. The muon counters are designed to 
count muons through a segmented detector. The segments of the Auger muon 
counters are scintillator bars whereas the segments of the AGASA muon counters 
were proportional counters. The limitation to measure a given number of incident 
muons is given by the number of segments of the counters.  

In general, the principle of operation of the muon counters is based on a 
binary logic in which each channel of the electronics, associated to a given 
segment of the detector, is able to differentiate between a state in which 
the signal is larger than a given threshold level and the one in which it is 
smaller. The threshold level is chosen in such a way that almost all muons 
can be identified, i.e.~the efficiency of each segment is close to $100\, \%$. 
For the case in which the signal is larger than the threshold level, the 
segment is said to be \emph{on} and otherwise \emph{off}. The time structure 
of the signal corresponding to one muon limits the time interval in which it 
is possible to identify single muons. This leads to the definition of a time 
interval, usually called inhibition window, in which it is decided whether 
a given segment is \emph{on} or \emph{off}. As a consequence, if one or more 
muons hit the same segment in a time interval of the order of the one 
corresponding to the inhibition window, the segment is tagged as \emph{on}, 
losing the information about the number of muons that hit that segment. 
Therefore, when a given number of muons hit a muon counter in a time interval 
of the order of the one corresponding to the inhibition window, a number $k$ 
of segments \emph{on} is obtained. If the number of incident muons is much 
smaller than the number of segments, the random variable $k$ is close the 
the number of muons that hit the counter. However, if the number of incident 
muons is close to the number of segments, the variable $k$ becomes much smaller 
than the number of incident muons. This effect is known as pile up \cite{Supa:08}.       

In this work, we find an analytic expression for the distribution function 
of $k$, the number of segments \emph{on}, given the number of incident muons,
$n_\mu$. In this case, $k$ is the random variable and $n_\mu$ is taken as a 
parameter of the distribution. The expressions for the mean value and the 
variance of $k$ are inferred by using the new formula, these expressions are 
equal to the ones obtained in Ref.~\cite{Supa:08} by using a different approach.
We also study how to estimate the parameter $n_\mu$ from measured values of 
$k$ and how to obtain a confidence interval. These studies are done for 192 
segments, which correspond to the total number of segments of the Auger muon 
counters and 50 segments that correspond to the number of segments of muon 
counters used in AGASA. 

It is worth mentioning that the main purpose of the muon counters is the 
reconstruction of the muon lateral distribution function (MLDF), i.e.~the muon 
density as a function of the distance to the shower axis, which is proportional 
to the mean value of $n_\mu$. Even though the estimation of the mean value of 
$n_\mu$, which is studied in Ref.~\cite{Ravignani:15}, is closely related to 
the determination of the MLDF, the estimation of $n_\mu$ is also important for 
different types of applications. In particular, it is important for the method 
used to reconstruct the MLDF developed in \cite{Supa:08}, in which an estimator 
of the number of muon that hit a given muon counter is inserted in the Poisson 
likelihood that approximates the exact likelihood in a given range of $n_\mu$. 
Also, the estimation of $n_\mu$ is necessary to obtain the calibration curve of 
the integrator (a complementary acquisition mode that muon counter can have), 
which is given by the mapping of the number of incident muons into the integrated 
signal \cite{AMIGA:19}. The estimation of $n_\mu$ is also relevant for studies 
related to the signal fluctuations and systematic uncertainties performed by 
using twin muon counters \cite{AMIGA:15}.

\section{Distribution function of the number of segments \emph{on} and muon number estimation}
\label{DF}

The mean value of the number of muons, $n_\mu$, that hit a muon counter is given by,
\begin{equation}
\label{Lmean}
\lambda = A\, \rho_\mu \, \cos\theta,
\end{equation} 
where $A$ is the area of the muon counters, $\rho_\mu$ is the muon density 
at a given distance to the shower axis, and $\theta$ is the zenith angle of 
the shower. Since the muon counters sample the MLDF at a given position in 
the shower plane, the number of muons that hit a given muon counter is a random
variable that follows the Poisson distribution. The distribution of $k$ given 
$\lambda$ has been obtained in Ref.~\cite{Ravignani:15} and is given by
\begin{equation}
P(k|\, \lambda) = {n_s \choose k}\, \exp(-\lambda) \left[ \exp\left(-\lambda/n_s\right) - 1 \right]^k,
\label{Pkmu}
\end{equation}
where $n_s$ is the number of segments of the muon counter. Note that this 
distribution does not depend on the number of muons that hit the muon
counters, which means that $P(k|\, \lambda)$ corresponds to the marginalization 
of the joint distribution $P(k,n_\mu|\, \lambda)$ with respect to $n_\mu$.        

On the other hand, since all segments are equal, the probability distribution 
function corresponding to a given configuration of the number of muons that hit 
each segment in a time interval corresponding to one inhibition window is given 
by the multinomial distribution,
\begin{equation}
P(n_{1},\ldots,n_{n_{s}}) = \frac{n_{\mu}!}{n_{1}! \ldots n_{n_{s}}!}%
\ \left( \frac{1}{n_{s}} \right)^{n_{\mu}},
\label{MultiNom}
\end{equation}
where $n_i$ is the number of muons that hit the $i-$th segment. Here 
$\{ n_{1},\ldots,n_{n_{s}} \}$ are such that $\sum_{i=1}^{n_{s}} n_{i} = n_{\mu}$.

The total number of segments \emph{on} can be written in the following
way,
\begin{equation}
k=\sum^{n_{s}}_{i=1} \widetilde{\Theta}(n_{i}),
\label{kon}
\end{equation}
where $\widetilde{\Theta}(n)=0$ if $n=0$ and $\widetilde{\Theta}(n)=1$ if 
$n \geq 1$. Note that $k \in \mathbb{N}$.

The mean value and the variance of $k$ can be calculated by using Eq.~(\ref{MultiNom}),
as reported in Ref.~\cite{Supa:08}. The mean value and the variance of $k$ are given 
by (see \ref{AppA} for details of the derivation),
\begin{eqnarray}
\label{MeanK} 
\langle k \rangle (n_{\mu})\!\!\!\! &=&\!\!\!\!  n_{s} \left[ 1-\left(1-%
\frac{1}{n_{s}}\right)^{n_{\mu}}\right], \\
\textrm{Var}\left[k\right] (n_{\mu})\!\!\!\! &=& \!\!\!\! n_{s}%
\left(1-\frac{1}{n_{s}}\right)^{n_{\mu}}%
\left[1+(n_{s}-1) \left(1-\frac{1}{n_{s}-1}\right)^{n_{\mu}} \right. \nonumber \\
&&-n_{s}\left. \left( 1-\frac{1}{n_{s}}\right)^{n_{\mu}} \right].
\label{Vark}
\end{eqnarray}

Beyond the fact that the mean value and the variance of $k$ can be calculated from 
Eq.~(\ref{MultiNom}) in a relatively direct way, an analytic expression of the 
distribution function of $k$ given $n_\mu$, $P(k|\, n_\mu)$, is more difficult 
to obtain. However, it can be inferred in a quite straightforward way by using the 
distribution $P(k|\, \lambda)$ given in Eq.~(\ref{Pkmu}). As mentioned before, 
$P(k|\, \lambda)$ is obtained marginalizing $P(k,n_\mu|\, \lambda)$ with respect to 
$n_\mu$. Besides, $P(k,n_\mu|\, \lambda)$ is given by the product of $P(k|\, n_\mu)$ 
with the Poisson distribution. Therefore,
\begin{eqnarray}
P(k|\, \lambda)\!\!\! &=&\!\!\! \sum_{n_\mu =\, 0}^\infty P(k|\, n_\mu) \, \exp(-\lambda)\,%
\frac{\lambda^{n_\mu}}{n_\mu !}  \nonumber \\
&=& \!\!\! {n_s \choose k}\, \exp(-\lambda) \left[ \exp\left(-\lambda/n_s\right)%
- 1 \right]^k. 
\label{Pkmu2}
\end{eqnarray}    
After introducing the following expression in Eq.~(\ref{Pkmu2}),
\begin{eqnarray}
\left[ \exp\left(-\lambda/n_s\right)- 1 \right]^k\!\!\! &=& \!\!\! \sum_{j\, =\, 0}^{k} %
{k \choose j} \, (-1)^j \exp\left( \lambda\, (k-j)/n_s \right), \nonumber \\
\exp\left( \lambda\, (k-j)/n_s \right)\!\!\!&=&\!\!\! \sum_{n_\mu =\, 0}^\infty %
\frac{\lambda^{n_\mu}}{n_\mu !} \left( \frac{k-j}{n_s} \right)^{n_\mu},
\end{eqnarray}    
the form of $P(k|\, n_\mu)$ is obtained comparing the terms proportional to 
$\lambda^{n_\mu}/n_\mu !$ in both sides of Eq.~(\ref{Pkmu}), 
\begin{equation}
P(k|\, n_\mu)={n_s \choose k}\, S(n_\mu,k) \frac{k!}{n_s^{n_\mu}}.
\label{Pknmu}
\end{equation}
Here $S(n_\mu,k)$ is the Stirling number of second kind, which is given by
\begin{equation}
S(n_\mu,k) = \frac{1}{k!} \, \sum_{j\, =\, 0}^{k} {k \choose j} \, %
(-1)^j \, \left( k-j \right)^{n_\mu}.
\end{equation}

The random variable $k$ ranges from $1$ to $n_s$. In the case in which 
$n_\mu < n_s$ and $k>n_\mu$, the condition $P(k|\, n_\mu) = 0$ must be 
fulfilled. To see this, it is enough to prove that $S(n_\mu,k)=0$ 
when $n_\mu < n_s$ and $k>n_\mu$. For that purpose, let us consider the 
following expression of the Stirling number of second kind,
\begin{equation}
S(n_\mu,k) = \frac{(-1)^k}{k!} \, \sum_{i\, =\, 0}^{k} {k \choose i} \, %
(-1)^i \, i^{n_\mu}.
\end{equation} 
From this expression, it is easy to see that,
\begin{equation}
S(n_\mu,k) = \left. \hat{\mathcal{O}}^{(n_\mu)} f_k(x) \right|_{x=1},
\end{equation}    
where $\hat{\mathcal{O}}^{(n_\mu)}$ corresponds to the differential
operator $\hat{\mathcal{O}}= x\, d/dx$ applied $n_\mu$ times to the 
function $f_k(x)=(1-x)^k$. By using the definitions given above, the
following expression is obtained,
\begin{equation}
\hat{\mathcal{O}}^{(n_\mu)} f_k(x) = \sum_{i=1}^{n_\mu} a_i \, x^i \, \frac{d^i f_k}{d x^i}(x),
\label{Onmufk}
\end{equation}
where $a_i$ are constant numbers. If $k>n_\mu$, all derivatives in 
Eq.~(\ref{Onmufk}) are proportional to a non-null and positive power 
of $(1-x)$. Therefore, it follows that 
$\hat{\mathcal{O}}^{(n_\mu)} f_k(x)|_{x=1} = 0$, which proves that 
$S(n_\mu,k) = 0$ for $k>n_\mu$ and then $P(k|\, n_\mu)=0$ for $k>n_\mu$.

Since $P(k|\, n_\mu)$ is a distribution function, it has to be normalized.
To see that this is true for the expression in Eq.~(\ref{Pknmu}), let
us consider the following property of the Stirling numbers of second 
kind \cite{Stanley:12},
\begin{equation}
x^{n_\mu} =\sum_{k=1}^{n_\mu} S(n_\mu,k)\, (x)_k, 
\label{xnmu}
\end{equation} 
where $(x)_k = x(x-1)...(x-k+1)$ is the falling factorial. If $x=n_s$,
Eq.~(\ref{xnmu}) becomes,
\begin{equation}
n_s^{n_\mu} = \left\{ 
\begin{array}{ll}
{\mathop{\displaystyle \sum_{k=1}^{n_\mu} }} S(n_\mu,k) \, (n_s)_k  &  n_\mu \leq n_s  \\[0.4cm]
{\mathop{\displaystyle \sum_{k=1}^{n_s} }} S(n_\mu,k)\, (n_s)_k + %
{\mathop{\displaystyle \sum_{k=n_s+1}^{n_\mu} }} S(n_\mu,k)\, (n_s)_k & n_\mu > n_s
\end{array} \right..
\label{nsnmu}
\end{equation} 
Using that $S(n_\mu,k)=0$ for $k>n_\mu$ and that $(n_s)_k = 0$ for $k>n_s$, 
Eq.~(\ref{nsnmu}) can be written as
\begin{equation}
n_s^{n_\mu} =\sum_{k=1}^{n_s} {n_s \choose k} \, S(n_\mu,k) \, k!,  
\label{nsnmu2}
\end{equation} 
which implies that
\begin{equation}
\sum_{k=1}^{n_s} P(k|\, n_\mu) = \sum_{k=1}^{n_s} {n_s \choose k} \, S(n_\mu,k) \, %
\frac{k!}{n_s^{n_\mu}}=1.  
\label{nsnmu3}
\end{equation} 

It is worth mentioning that from Eq.~(\ref{Pknmu}) it is possible to calculate
the mean value and the variance of $k$ which have to be equal to the expressions 
obtained by using the multinomial distribution (see Eqs.~(\ref{MeanK}) and 
(\ref{Vark})). The mean value of $k$ is given by
\begin{equation}
\langle k \rangle = \sum_{k=1}^{n_s} {n_s \choose k} \, S(n_\mu,k)\, k \, %
\frac{k!}{n_s^{n_\mu}}.
\label{MeanKNew}
\end{equation}
From the recurrence satisfied by the Stirling numbers of second kind,
$S(n_\mu,k) = k\, S(n_\mu-1,k) + S(n_\mu-1,k-1)$ \cite{Stanley:12}, it 
follows that $k\, S(n_\mu,k) = S(n_\mu+1,k) + S(n_\mu,k-1)$. Introducing
this equality in Eq.~(\ref{MeanKNew}), the following expression is obtained,
%
\begin{equation}
\langle k \rangle = \sum_{k=1}^{n_s} {n_s \choose k} \, S(n_\mu+1,k)\, %
\frac{k!}{n_s^{n_\mu}} - %
 n_s \sum_{j=1}^{n_s-1} {n_s-1 \choose j} \, S(n_\mu,j)\, %
\frac{j!}{n_s^{n_\mu}},
\label{MeanKNew2}
\end{equation}
%
where the second term of this equation is obtained by doing the change of 
variable $j=k-1$. From Eq.~(\ref{nsnmu2}), it is easy to see that 
$\langle k \rangle = n_s^{n_\mu+1}/n_s^{n_\mu} - n_s \, (n_s-1)^{n_\mu}/n_s^{n_\mu}$,
which after some algebra becomes  
$\langle k \rangle = n_s\, (1 - (1-1/n_s)^{n_\mu})$, i.e.~the same expression 
obtained by using the multinomial distribution. In a similar way, but in this 
case using the recurrence relation satisfied by the Stirling number of second 
kind twice, it is possible to calculate the variance of $k$, whose expression 
obtained in this way is the same as the one obtained by using the multinomial 
distribution. 

\begin{figure}[ht!]
\centering
\includegraphics[width=9cm]{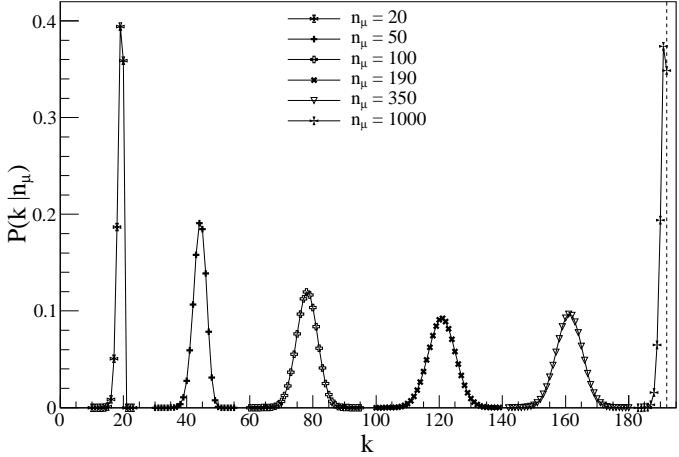}
\caption{$P(k|\, n_\mu)$ as a function of $k$ for $n_s=192$. The vertical dashed line
corresponds to $k=192$. The solid lines joining the discrete points are added to guide 
the eye. \label{Dist}}
\end{figure}

The Auger muon counters installed in each position of the array are composed of 
three modules of 64 segments each summing a total of 192 segments (see 
Ref.~\cite{AMIGA:19} for details). Figure \ref{Dist} shows $P(k|\, n_\mu)$ as a 
function of $k$ for different values of $n_\mu$. As expected, the maximum of the 
distribution is shifted towards larger values of $k$ for increasing values of 
$n_\mu$. Note that even for $n_\mu=1000$ the distribution $P(k|\, n_\mu)$ presents 
a maximum. 

\begin{figure}[ht!]
\centering
\includegraphics[width=9cm]{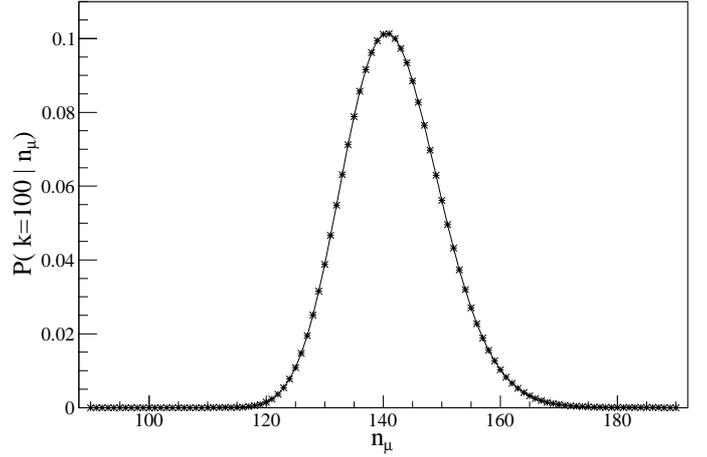}
\includegraphics[width=9cm]{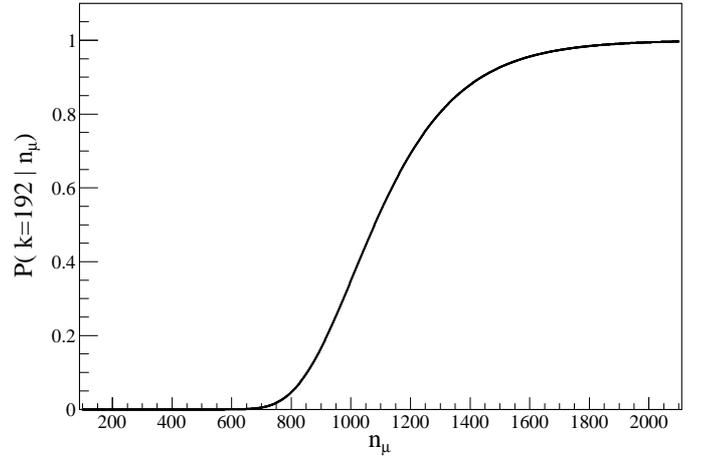}
\caption{Likelihood function $L(n_\mu)=P(k|\, n_\mu)$ as a function of $n_\mu$ for $n_s=192$
and for $k=100$ (top panel) and $k=192$ (bottom panel). The solid lines joining the discrete 
points are added to guide the eye.\label{BFex}}
\end{figure}

As a result of a measurement, a given value of $k$ is obtained. From this value of
$k$ it is possible to estimate the parameter $n_\mu$ and to determine a confidence
interval at a given confidence level. The maximum likelihood estimator of $n_\mu$,
$\hat{n}_\mu$, is obtained by finding the maximum of the likelihood function 
$L(n_\mu)=P(k|\, n_\mu)$, i.e.,
\begin{equation}
\hat{n}_\mu=\underset{n_\mu \in \mathbb{N}} {\arg\max}\, P(k|\, n_\mu).
\label{Eqnhat}
\end{equation}
In this case, $\hat{n}_\mu$ has to be calculated numerically for each particular value
of $k$ measured. Figure \ref{BFex} shows the likelihood function as a function of 
$n_\mu$ for $k=100$ and $k=192$. From the top panel of the figure, it can be seen that 
for $k=100$ the likelihood presents a well-defined maximum as for all other allowed 
values of $k$ except for $k=192$. In this case, as can be seen from the bottom panel 
of the figure, the likelihood reaches a maximum for $n_{\mu} \rightarrow \infty$. This 
means that when the value $k=192$ is obtained as a result of a measurement, only a 
lower limit of $n_\mu$ can be found, as shown below. Note that 
$P(k=n_s|\, n_\mu) \rightarrow 1$ when $n_{\mu} \rightarrow \infty$.

\begin{figure}[ht!]
\centering
\includegraphics[width=9cm]{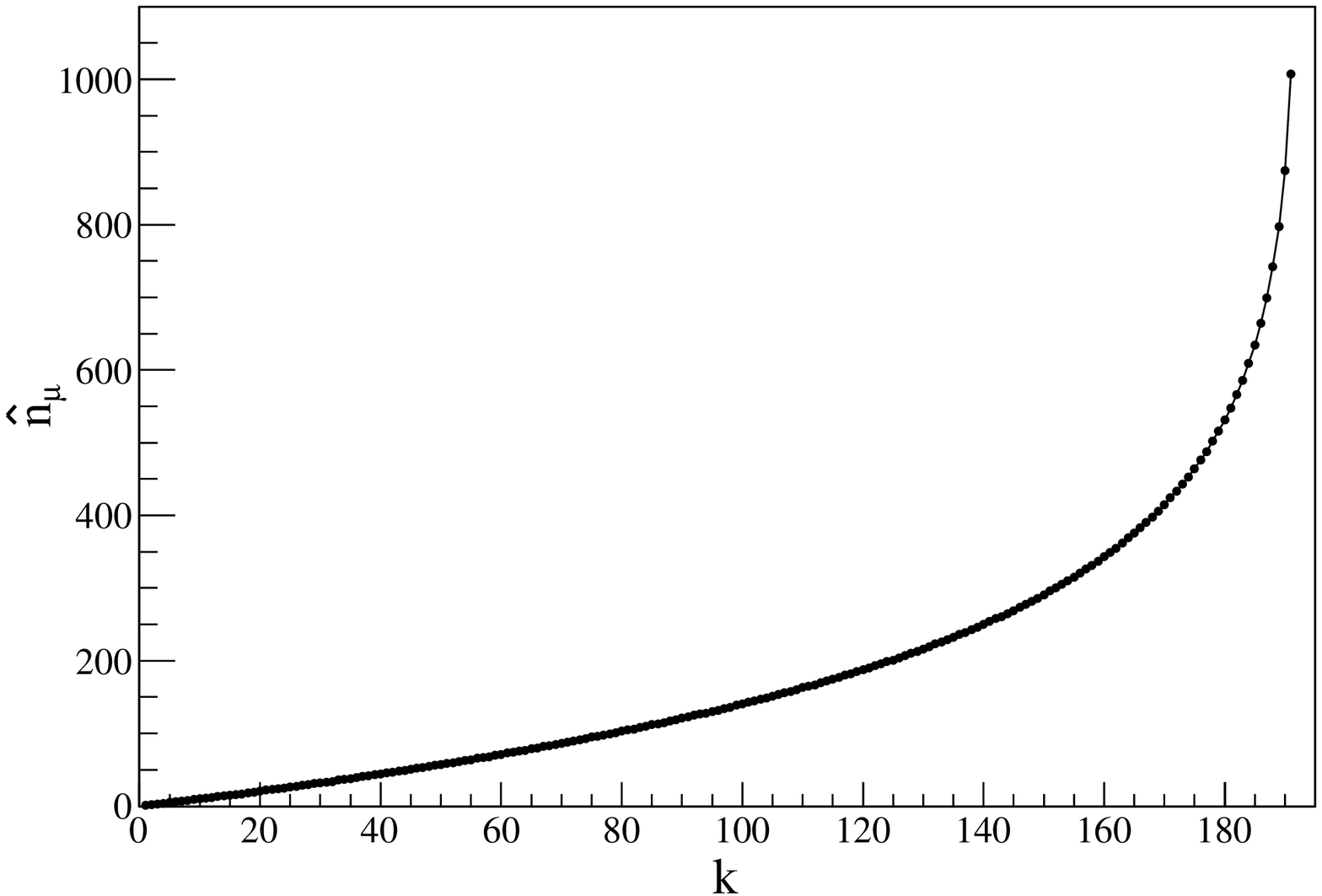}
\includegraphics[width=9cm]{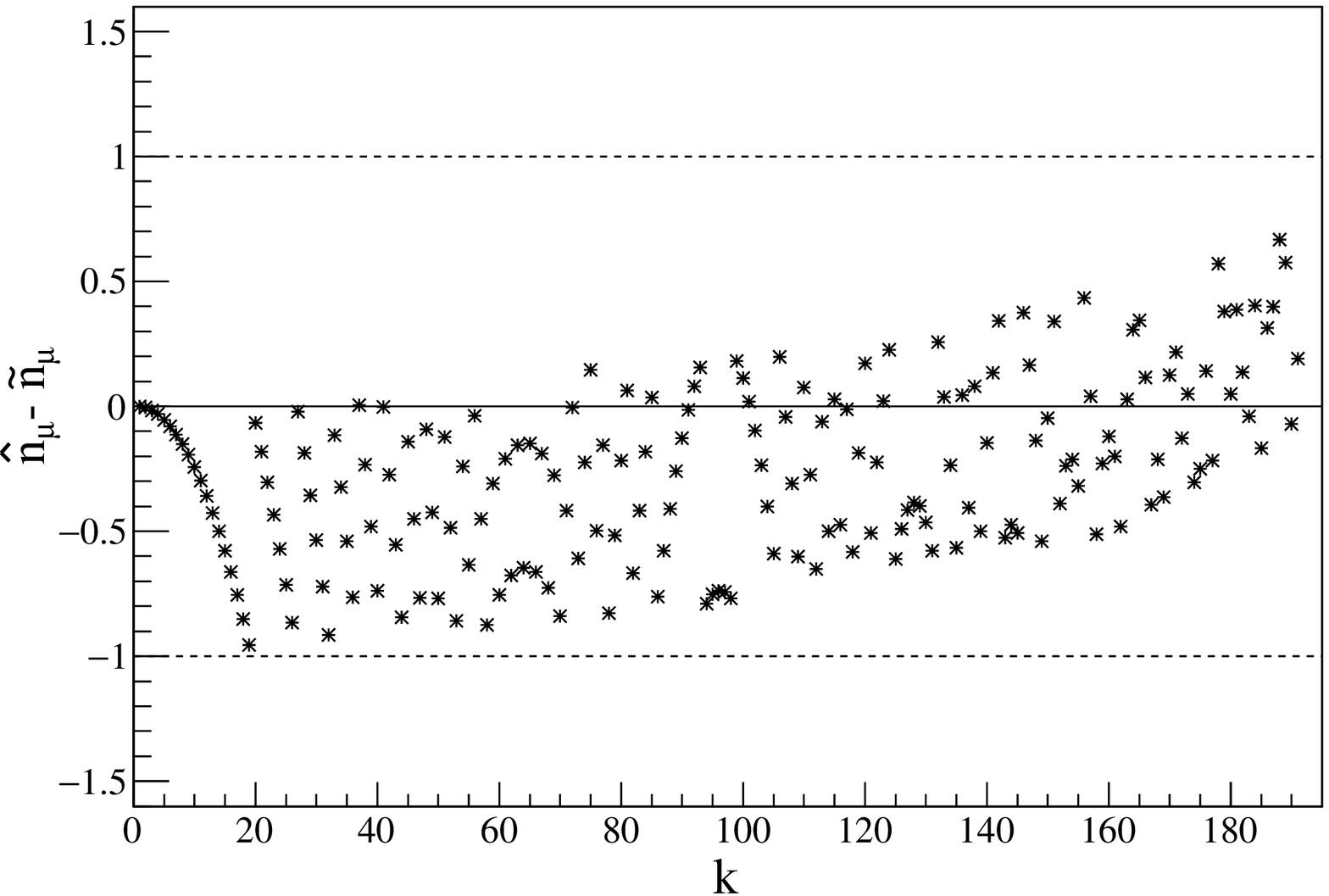}
\caption{Top panel: Maximum likelihood estimator of $n_\mu$, $\hat{n}_\mu$, as a function of $k$. 
The solid line joining the discrete points is added to guide the eye. Bottom panel: Difference 
between $\hat{n}_\mu$ and $\widetilde{n}_\mu$ (approximated expression of $\hat{n}_\mu$ given by
Eq.~(\ref{ntilde})) as a function of $k$. The number of segments considered is $n_s=192$. 
\label{BF}}
\end{figure}

Figure \ref{BF} shows $\hat{n}_\mu$ as a function of $k$ obtained by maximizing the likelihood
function for $n_s=192$. From the plot, it can be seen that $\hat{n}_\mu \cong k$ for small values 
of $k$. Moreover, $\hat{n}_\mu$ starts to deviate from $k$ in more than $10\, \%$ at $k \cong 40$. 
For larger values of $k$, $\hat{n}_\mu$ starts to increase faster in such a way 
that $\hat{n}_\mu = 1007$ for $k=191$.

As proposed in Ref.~\cite{Supa:08}, a good approximation of $\hat{n}_\mu$ 
can be found by using the expression corresponding to the mean value of $k$ given in 
Eq.~(\ref{MeanK}). Inverting Eq.~(\ref{MeanK}), the expression 
\begin{equation}
\widetilde{n}_\mu = \frac{\ln\left(1-{\mathop{\displaystyle \frac{k}{n_s} }} \right)}%
{\ln\left(1-{\mathop{\displaystyle{\frac{1}{n_s}} }} \right) }, 
\label{ntilde}
\end{equation}    
is obtained, where the mean value of $k$ is replaced by the random variable $k$. The 
bottom panel of Fig.~\ref{BF} shows the difference between the maximum likelihood 
estimator of $n_\mu$, $\hat{n}_\mu$, and the approximated expression of Eq.~(\ref{ntilde}). 
It can be seen that $\widetilde{n}_\mu$ differs in less than one from $\hat{n}_\mu$ in 
the whole range of the variable $k$, which shows that $\widetilde{n}_\mu$ is a very good 
approximation of $\hat{n}_\mu$.  

\begin{figure}[ht!]
\centering
\includegraphics[width=9cm]{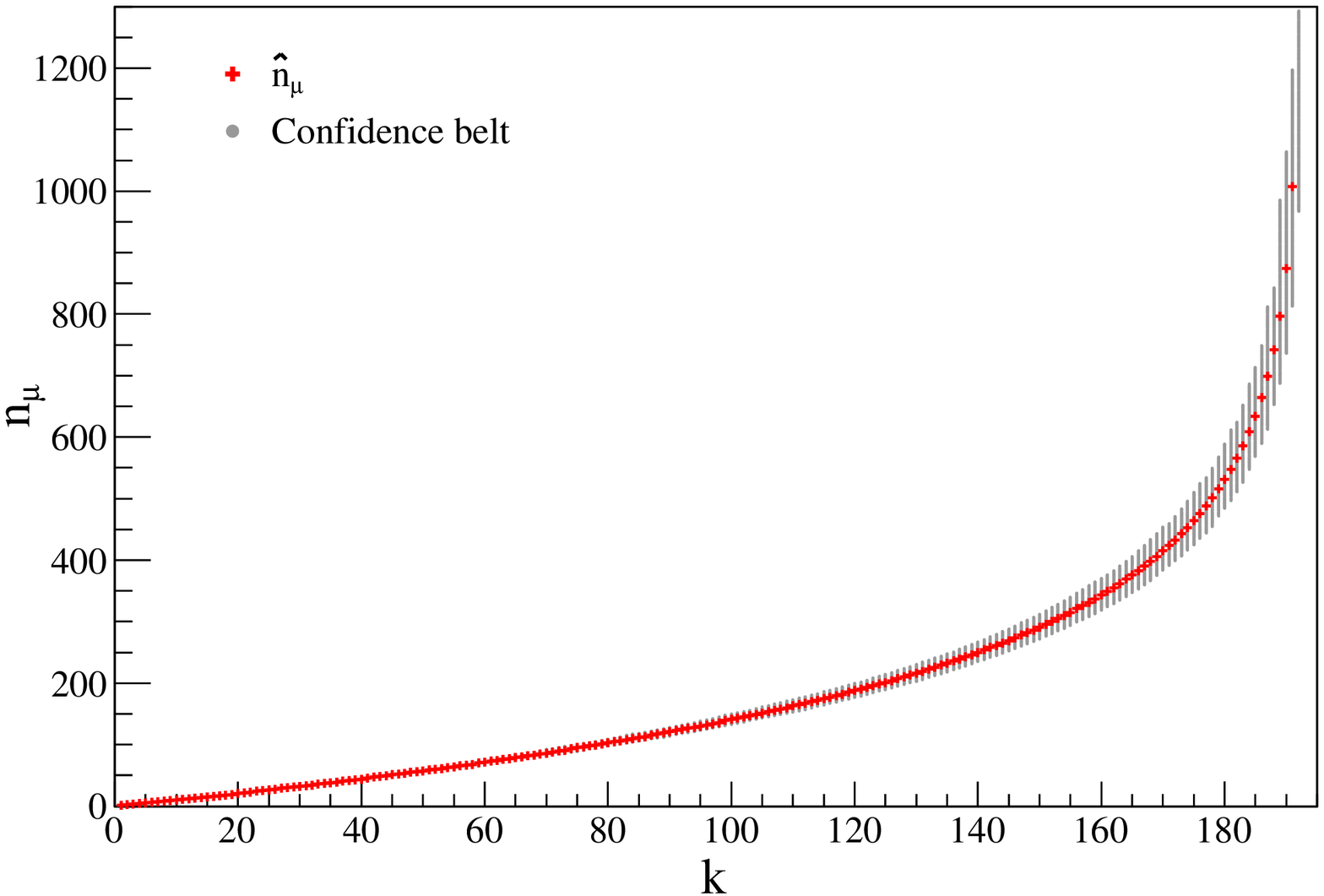}
\includegraphics[width=9cm]{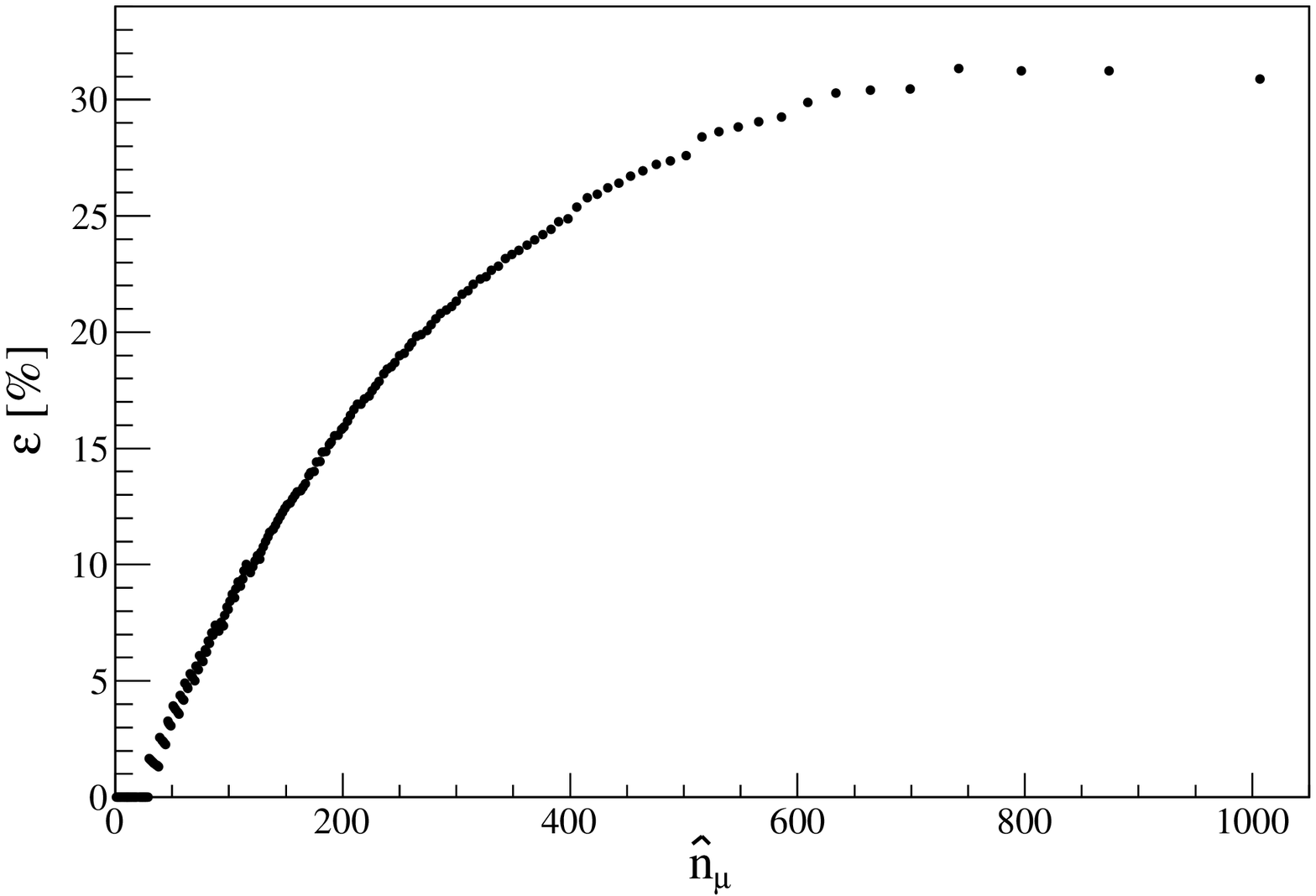}
\caption{Top panel: Confidence belt of $P(k|\, n_\mu)$ and $\hat{n}_\mu$ as a function of
$k$. Bottom panel: $\varepsilon$ as a function of $\hat{n}_\mu$. The number of segments 
considered is $n_s=192$ and the confidence level used in the calculation is $1-\alpha=0.6827$. 
\label{CLBF}}
\end{figure}

The confidence belt of the distribution function $P(k|\, n_\mu)$ is constructed by 
finding the values of $k$, for a given $n_\mu$, that satisfy
\begin{equation}
\sum_k P(k|\, n_\mu) \geq 1-\alpha,
\end{equation}   
where $1-\alpha$ is the confidence level (CL). In this work, the ordering proposed
by Feldman and Cousins \cite{Feldman:98} is considered for the calculation of the 
confidence belt. The top panel of Fig.~\ref{CLBF} shows the confidence belt of 
$P(k|\, n_\mu)$ for $n_s=192$ and $1-\alpha=0.6827$. The $\hat{n}_\mu$ as a function 
of $k$ is also shown. From the figure it can be seen that the width of the confidence 
belt increases with $k$, as expected. It can also be seen that for $k=n_s=192$ only a
lower limit can be obtained since the case $k=n_s$, i.e., all segments of the counter 
\emph{on} is compatible with a semi-infinite set of $n_\mu$ values at any confidence
level.   

The relative uncertainty of $\hat{n}_\mu$ is defined here as
\begin{equation}
\label{eps}
\varepsilon = \frac{n_\mu^{max} - n_\mu^{min}}{2\, \hat{n}_\mu},
\end{equation}
where $n_\mu^{max}$ and $n_\mu^{min}$ are the maximum and the minimum values
of $n_\mu$, respectively, for a given value of $k$ obtained at a given CL. The 
bottom panel of Fig.~\ref{CLBF} shows $\varepsilon$ as a function of $\hat{n}_\mu$. 
As expected, $\varepsilon$ increases with $\hat{n}_\mu$ taking values smaller 
than $\sim 16\, \%$ for $\hat{n}_\mu \lesssim 201$ and reaching values of the order 
of $30\, \%$ in the region where $700 \lesssim \hat{n}_\mu \lesssim 900$ (close 
to $k=192$). 

\begin{figure}[ht!]
\centering
\includegraphics[width=9cm]{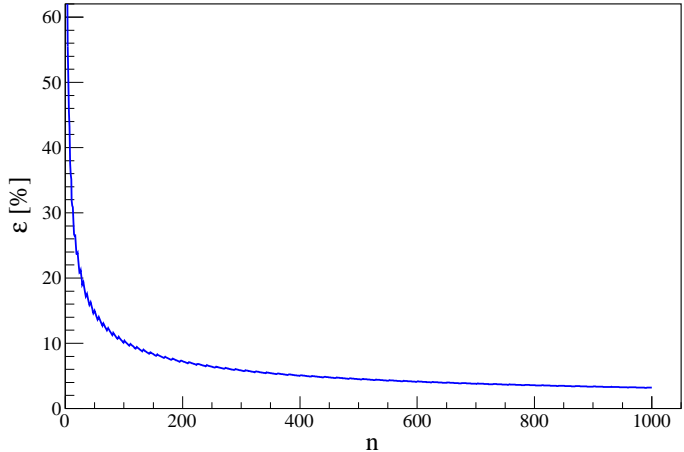}
\caption{Relative uncertainty corresponding to the maximum likelihood estimator of the 
mean value of the Poisson distribution as a function of $n$, a measured value of a Poisson 
random variable. The confidence level used in the calculation is $1-\alpha=0.6827$. 
\label{EpsPoiss}}
\end{figure}

The determination of the mean value of the number of muons at a given distance to 
the shower axis $\lambda$, given by Eq.~(\ref{Lmean}), is affected by the segmentation 
of the detector in combination with the pile-up effect and also by the Poisson 
fluctuations. Figure \ref{EpsPoiss} shows the relative uncertainty corresponding to the 
estimator of the mean value of the Poisson distribution as a function of $n$, a measured 
valued of a Poisson random variable, which in this case coincides with the maximum likelihood 
estimator of the mean value. This relative uncertainty is obtained following the same 
procedure used to calculate the relative uncertainty corresponding to the $\hat{n}_\mu$ 
estimator. Comparing the bottom panel of Fig.~\ref{CLBF} with Fig.~\ref{EpsPoiss} it can 
be seen that for values of $\hat{n}_\mu$ smaller than $\sim 113$ the uncertainty on the 
determination of the mean value of the number of muons is dominated by the Poisson 
fluctuations but for values of $\hat{n}_\mu$ larger than $\sim 113$ the dominant uncertainty 
is the one introduced by the segmentation of the detector in combination with the pile-up 
effect.   
%

\begin{figure}[ht!]
\centering
\includegraphics[width=9cm]{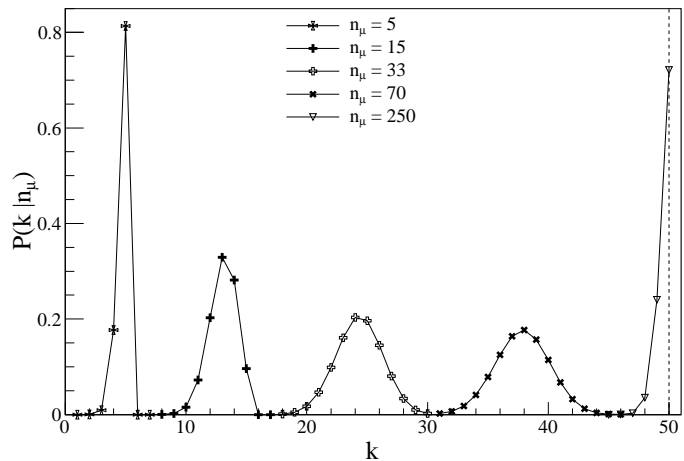}
\caption{$P(k|\, n_\mu)$ as a function of $k$ for $n_s=50$. The vertical dashed line
corresponds to $k=50$. The solid lines joining the discrete points are added to guide 
the eye. \label{DistNs50}}
\end{figure}

The AGASA experiment used muon counters of 50 segments to measure the muon content
of the showers. Figure \ref{DistNs50} shows the distribution function $P(k|\, n_\mu)$
as a function of $k$ for different values of $n_\mu$ and for $n_s=50$. As expected,
for $n_\mu=250$ the maximum of the distribution is reached at $k=50$, which means
that these counters saturate with a much smaller number of incident muons compared 
to the ones with 192 segments.

\begin{figure}[ht!]
\centering
\includegraphics[width=9cm]{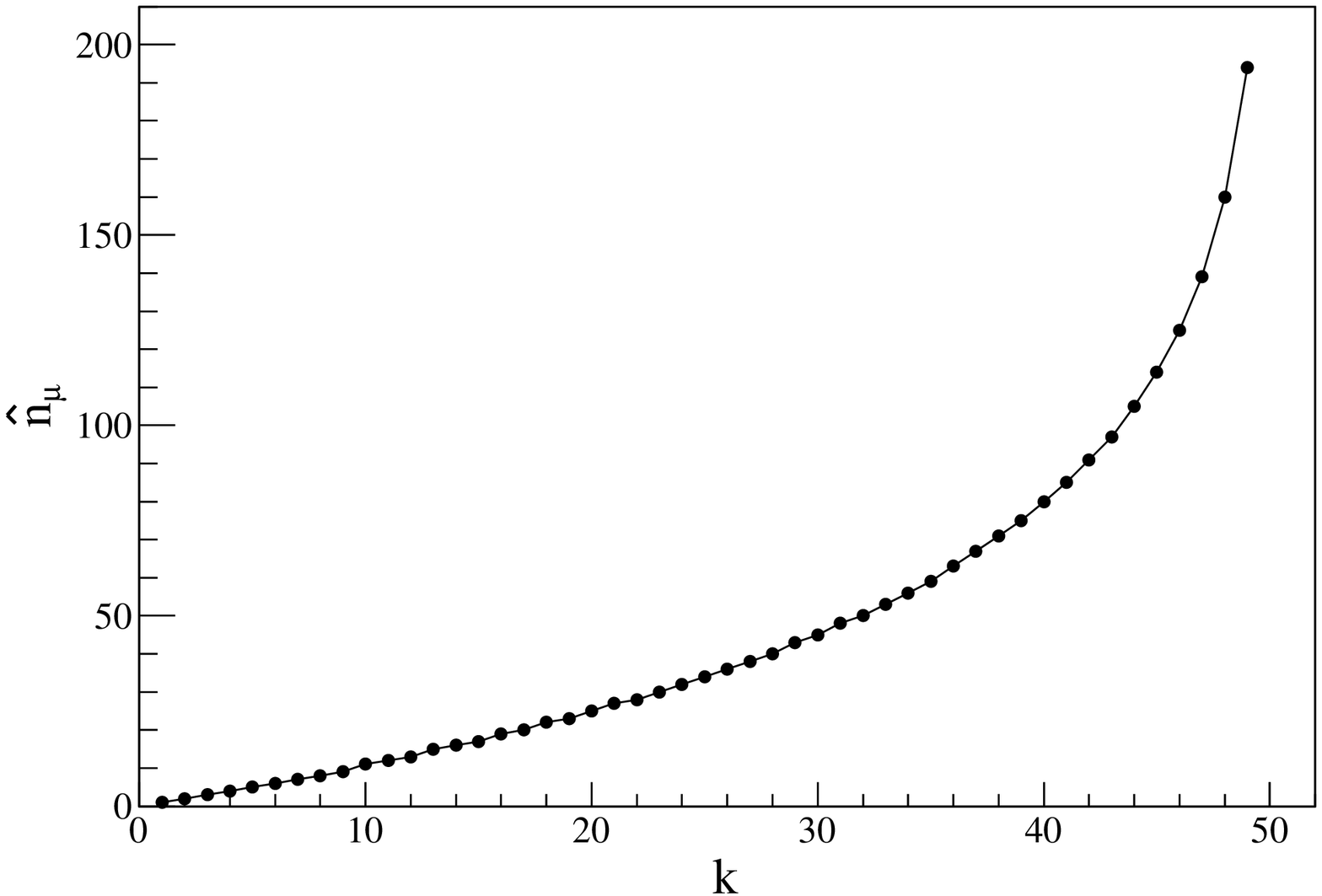}
\includegraphics[width=9cm]{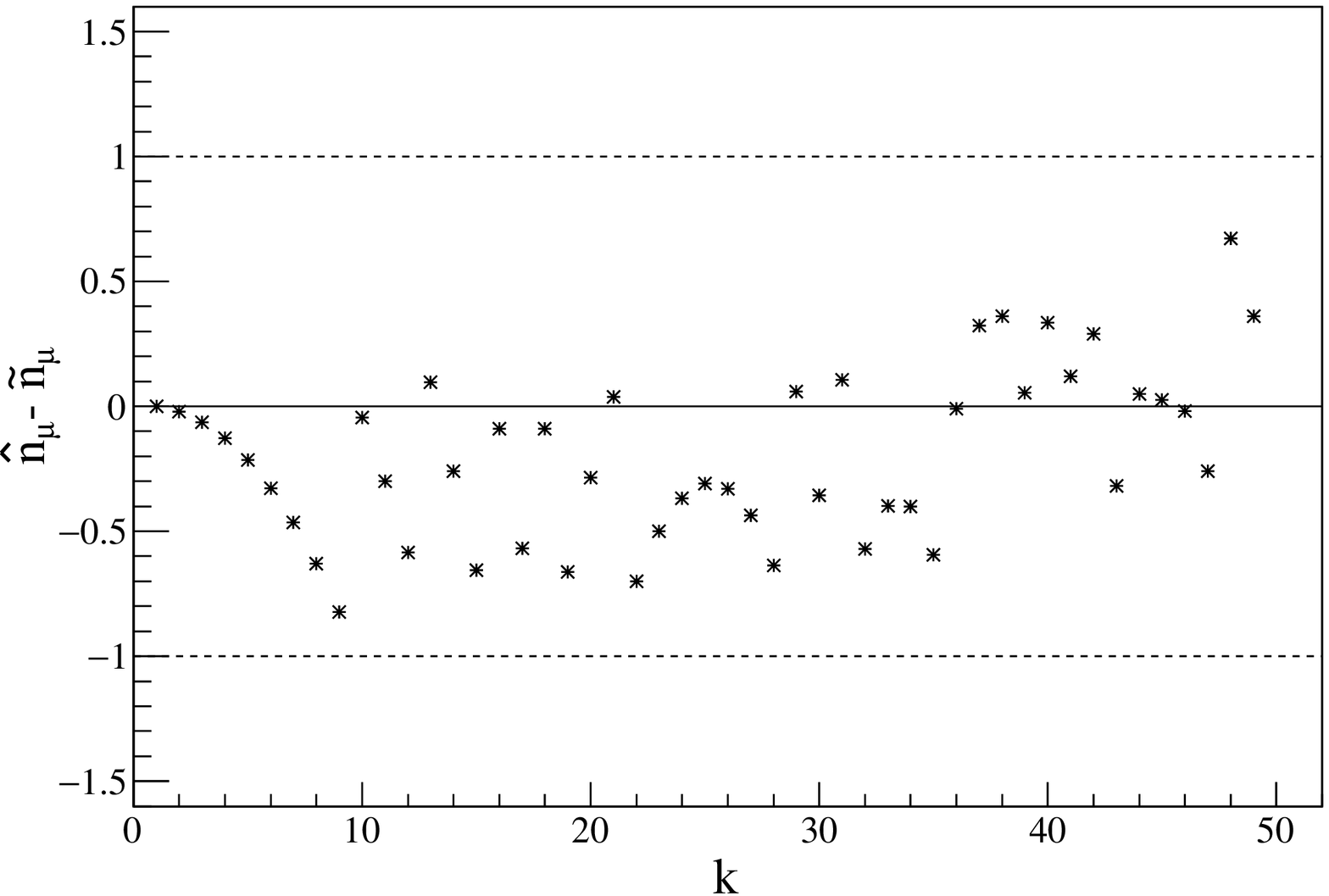}
\caption{Top panel: Maximum likelihood estimator of $n_\mu$, $\hat{n}_\mu$, as a function of $k$.
The solid line joining the discrete points is added to guide the eye. Bottom panel: Difference 
between $\hat{n}_\mu$ and $\widetilde{n}_\mu$ (approximated expression of $\hat{n}_\mu$ given in
Eq.~(\ref{ntilde})) as a function of $k$. The number of segments considered is $n_s=50$.
\label{BFNs50}}
\end{figure}

The top panel of Fig.~\ref{BFNs50} shows the maximum likelihood estimator of
$n_\mu$ calculated numerically by using Eq.~(\ref{Eqnhat}) for $n_s=50$. As for
the $n_s=192$ case, $\hat{n}_\mu \cong k$ for small values of $k$ in such a way
that for $k>20$, the departure of $\hat{n}_\mu$ from $\hat{n}_\mu \cong k$ becomes 
larger than $10\, \%$. For larger values of $k$, $\hat{n}_\mu$ starts to increase 
faster in such a way that $\hat{n}_\mu = 194$ for $k=49$.

The bottom panel of Fig.~\ref{BFNs50} shows $\hat{n}_\mu - \widetilde{n}_\mu$ as a function
of $k$. Also in this case, the absolute value of this difference is smaller than one, which 
indicates that $\widetilde{n}_\mu$, given in Eq.~(\ref{ntilde}), is a very good approximation 
of $\hat{n}_\mu$ also for the $n_s = 50$ case. 

\begin{figure}[ht!]
\centering
\includegraphics[width=9cm]{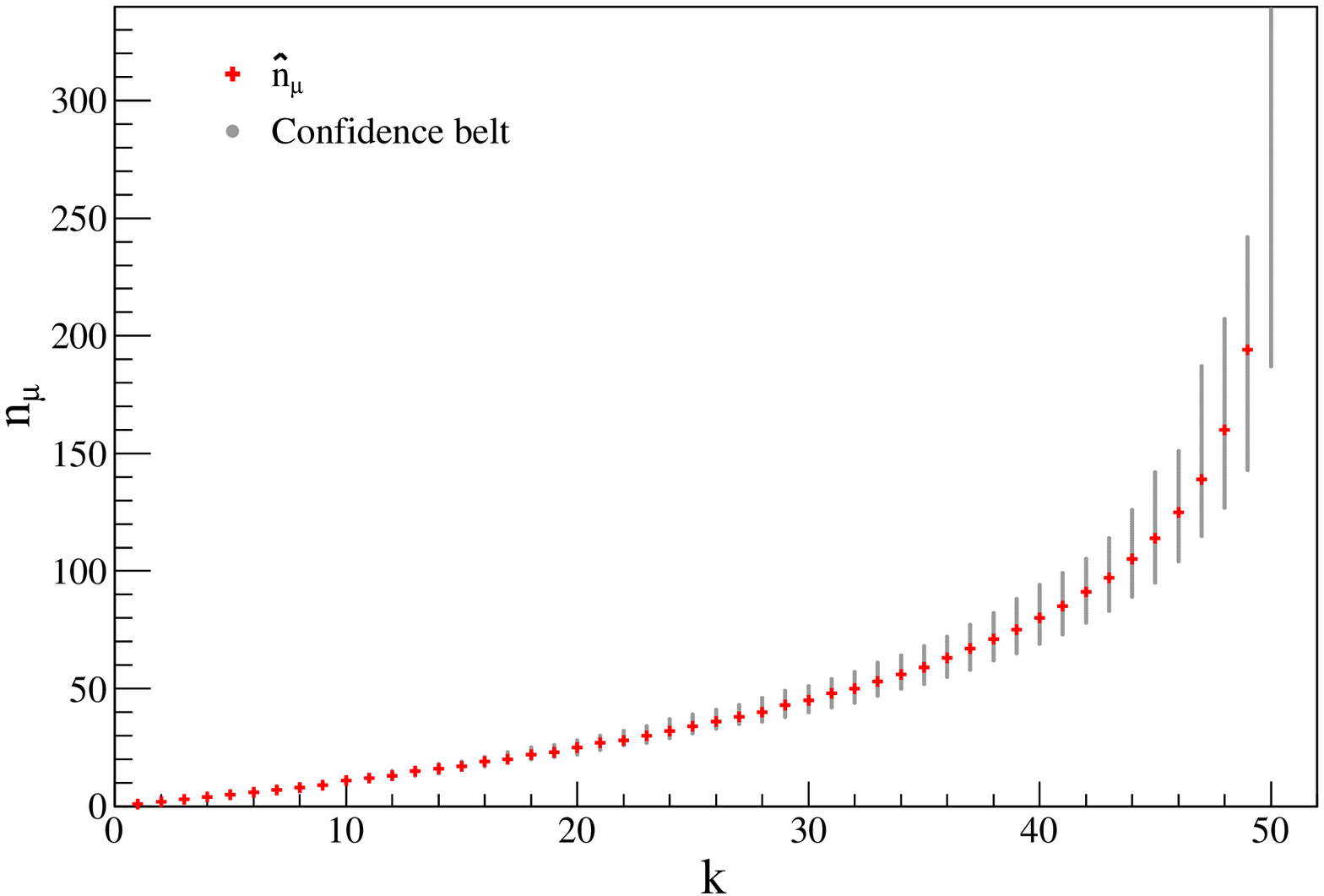}
\includegraphics[width=9cm]{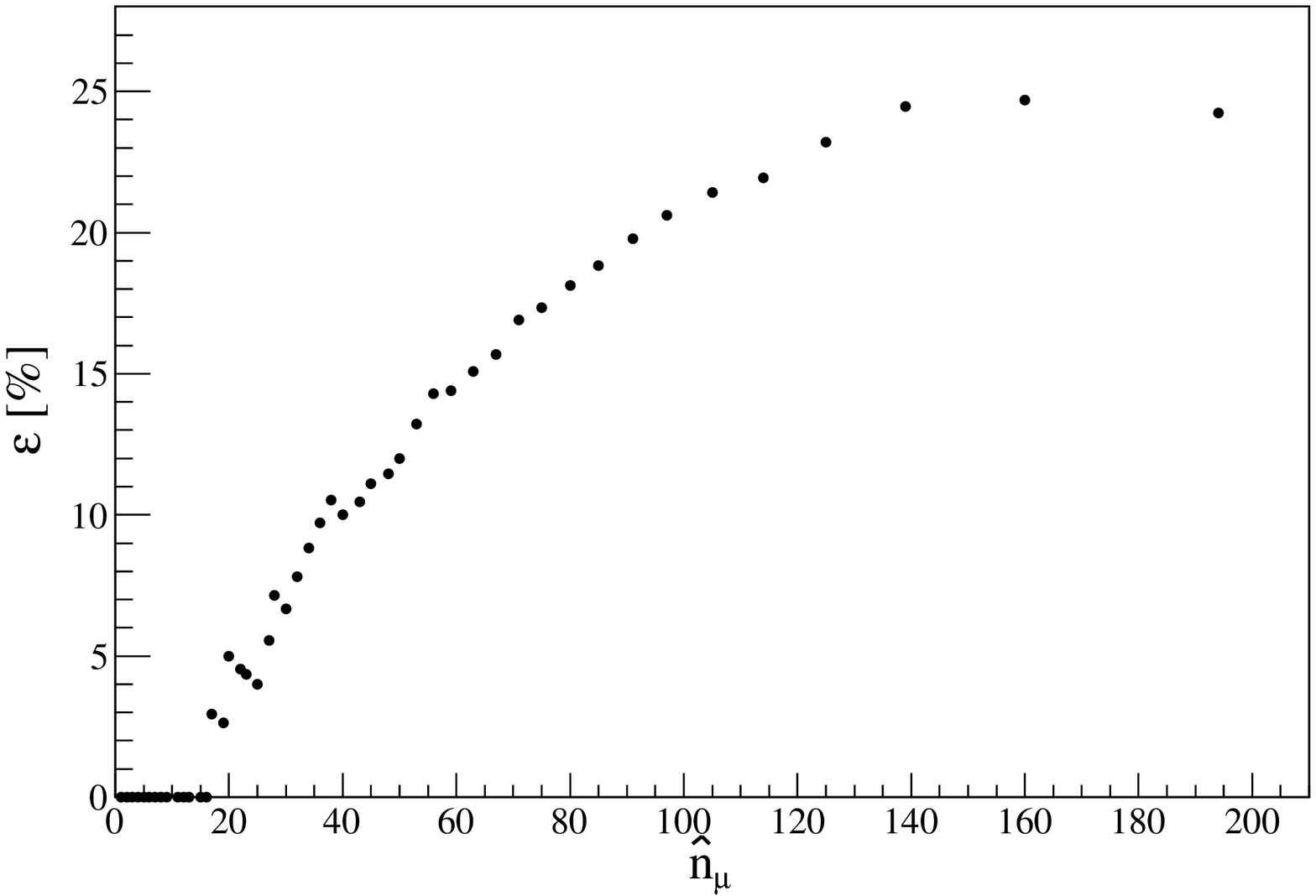}
\caption{Top panel: Confidence belt of $P(k|\, n_\mu)$ and $\hat{n}_\mu$ as a function of
$k$. Bottom panel: $\varepsilon$ as a function of $\hat{n}_\mu$. The number of segments 
considered is $n_s=50$ and the confidence level used in the calculation is $1-\alpha=0.6827$.
\label{CLBFNs50}}
\end{figure}

The top panel of Fig.~\ref{CLBFNs50} shows the confidence belt of $P(k|\, n_\mu)$ for $n_s = 50$
and $1-\alpha=0.6827$. The $\hat{n}_\mu$ as a function of $k$ is also shown in the plot. As in the 
case corresponding to $n_s = 192$, the confidence belt becomes wider for increasing values of
$k$. This behavior becomes more evident from the plot in the bottom panel of the same figure,
in which it can be seen that the relative uncertainty of $\hat{n}_\mu$ is smaller than 
$\sim 16\, \%$ for $k \leq 67$ and takes values close to $24\, \%$ in the region where 
$140 \lesssim k \lesssim 194$. 

As in the previous case, comparing the bottom panel of Fig.~\ref{CLBFNs50} with 
Fig.~\ref{EpsPoiss} it can be seen that for values of $\hat{n}_\mu$ smaller than $\sim 55$
the uncertainty on the determination of the mean value of the number of muons is dominated 
by the Poisson fluctuations but for values of $\hat{n}_\mu$ larger than $\sim 55$ the dominant 
uncertainty is the one introduced by the segmentation of the detector in combination with the 
pile-up effect.

Note that the relative uncertainty on the determination of $\hat{n}_\mu$ for the AGASA muon
counters cannot be compared straightforwardly with the one corresponding to Auger, since the 
area of the AGASA muon counters is 25 m$^2$ whereas the one corresponding to the Auger muon 
counters is 30 m$^2$. Therefore, the number of muons that hit an AGASA muon detector is, on 
average, $\sim 17\, \%$ smaller than the one corresponding to an Auger muon detector, provided 
that the muon flux that hit the detectors is the same. In any case, the determination of the 
muon density at a given distance to the shower axis done by using the Auger muon detectors 
should be better than the one corresponding to the AGASA muon detectors, since the Auger muon 
detectors have a larger area and a larger number of segments ($A = 30$ m$^2$ and $n_s = 192$) 
than the ones corresponding to AGASA ($A = 25$ m$^2$ and $n_s = 50$). 

The estimation of the number of muons studied in this work corresponds to ideal muon detectors. 
Real detectors are subject to different effects that have to be taken into account in order 
not to introduce biases in the estimated number of muons. For instance, there are three main 
effects that can introduce biases in the Auger muon detectors. The first one is the noise 
produced by the dark rate of the silicon photomultipliers, which can reach the discriminator 
threshold due to the inner-cells crosstalk. This effect can be mostly reduced choosing a proper 
counting strategy (see Ref.~\cite{AMIGA:19} for details). The second one is the efficiency of 
each segment, which can be smaller than 100 \%. In this case a correction can be obtained 
from the estimated efficiency, which is measured in the laboratory \cite{AMIGA:19}. Note that 
the efficiency of the Auger muon detectors is $\sim 98.5$ \%. The third one is caused by particles 
passing through two adjacent segments, the bias introduced by this effect can be estimated 
form detailed simulations of the detector \cite{Figueira:17}. Note that the sources of biases 
on the estimation of the number of incident muons depend on the specific design of the muon 
detector under consideration and have to be studied in detail for each particular case.

\section{Conclusions}

In this work we have studied in detail the estimation of the number of muons
that hit a muon counter from the number of segments \emph{on}, $k$, which is 
the random variable that is measured in an experiment. For that purpose we have 
found an analytic expression for the distribution function of $k$, given a number 
of incident muons. We have considered the number of segments corresponding to the 
muon counters of Auger and also the one corresponding to the muon counters of AGASA. 

We have found that for small values of $k$, compared with the number of segments, 
the estimator of the muon number is close to $k$ but increases much faster for 
larger values of $k$. We have also found that the relative uncertainty in the 
determination of the number of muons is small for small values of $k$ and that 
it increases relatively fast with $k$ reaching values close to $24$ and $30\, \%$,
for $n_s=50$ and $n_s=192$ respectively, in the region where $k$ is close to 
the total number of segments.   

The main motivation of these studies is the measurement of the muon content of 
air showers initiated by cosmic rays, which is intimately related to the chemical 
composition of the primary particle, an open problem of the high-energy 
astrophysics. However, it is worth mentioning that the methods developed in 
this work can be relevant in other applications.

\appendix
\section{Calculation of $\langle k \rangle$ and Var[$k$] from the multinomial distribution}
\label{AppA}

In this section, the steps that lead to the expressions for $\langle k \rangle$ and 
Var[$k$] (see Eqs.~(\ref{MeanK}) and (\ref{Vark})), obtained in Ref.~\cite{Supa:08} 
by using the multinomial distribution are given.   

Let us start with the calculation of the mean value of $k$. From Eq.~(\ref{kon}) it
can be seen that,
\begin{equation}
\langle k \rangle=\left\langle \sum^{n_{s}}_{i=1} \widetilde{\Theta}(n_{i}) \right\rangle = %
n_s \left\langle  \widetilde{\Theta}(n_{1}) \right\rangle.
\label{konapp}
\end{equation}
The distribution function of $n_1$ is given by the binomial distribution, i.e.
\begin{equation}
P_1(n_1) = {n_\mu \choose n_1} \, \left(\frac{1}{n_s} \right)^{n_1} \, %
\left(1 - \frac{1}{n_s} \right)^{n_\mu-n_1}.
\label{P1n1}
\end{equation}
Therefore,
\begin{equation}
\langle k \rangle = n_s \sum_{n_1=0}^{n_\mu} \widetilde{\Theta}(n_{1}) \, P_1(n_1) = 
n_s \sum_{n_1=1}^{n_\mu} P_1(n_1) = n_s \, (1-P_1(0)).
\label{MeanKapp}
\end{equation}
Combining Eqs.~(\ref{P1n1}) and (\ref{MeanKapp}), the expression for the mean value of $k$
given by Eq.~(\ref{MeanK}) is obtained.

The variance of $k$ is calculated in a similar way. For that purpose, let us first 
calculate the mean value of $k^2$, which is given by
\begin{eqnarray}
\langle k^2 \rangle \!\!\!&=& \!\!\!\left\langle \sum^{n_{s}}_{i=1}  \sum^{n_{s}}_{j=1} %
\widetilde{\Theta}(n_{i}) \, \widetilde{\Theta}(n_{j}) \right\rangle  \nonumber \\ %
\!\!\! &=& \!\!\! n_s \left\langle  \widetilde{\Theta}^2(n_{1}) \right\rangle + n_s (n_s-1) 
\left\langle \widetilde{\Theta}(n_{1}) \, \widetilde{\Theta}(n_{2}) \right\rangle.
\label{k2app}
\end{eqnarray}
To calculate the averages in Eq.~(\ref{k2app}) besides $P_1(n_1)$, $P_2(n_1,n_2)$ is required.
From Eq.~(\ref{MultiNom}) it can be seen that
\begin{equation}
P_2(n_1,n_2) = \frac{n_\mu!}{n_1! n_2! (n_\mu - n_1-n_2)!} \left(\frac{1}{n_s} \right)^{n_1+n_2} \, %
\left(1 - \frac{2}{n_s} \right)^{n_\mu-n_1-n_2},
\end{equation}
where $n_1+n_2 \leq n_\mu$. In a similar way to the one followed to obtain Eq.~(\ref{MeanKapp}),
the next expression for the mean value $k^2$ is obtained
\begin{eqnarray}
\langle k^2 \rangle \!\!\!&=& \!\!\! n_s \, (1-P_1(0)) + n_s (n_s-1)\times \nonumber \\
&& \!\!\! \left(1-P_2(0,0)-2 \sum_{n_1=1}^{n_\mu} P_2(n_1,0) \right).
\end{eqnarray}
By using that,
\begin{equation}
\sum_{n_1=1}^{n_\mu} P_2(n_1,0) = \left( 1 - \frac{1}{n_s} \right)^{n_\mu} - 
\left( 1 - \frac{2}{n_s} \right)^{n_\mu},
\end{equation}
the following expression is obtained,
\begin{equation}
\langle k^2 \rangle = n_s^2 - n_s (2 n_s-1) \left( 1 - \frac{1}{n_s} \right)^{n_\mu} +%
n_s (n_s-1) \left( 1 - \frac{2}{n_s} \right)^{n_\mu}.
\label{k2appF}
\end{equation}
From Eqs.~(\ref{k2appF}) and (\ref{MeanK}), the expression for the variance of $k$ given
by Eq.~(\ref{Vark}) is obtained.

\section*{Acknowledgements}

A.~D.~S.~is member of the Carrera del Investigador Cient\'ifico of CONICET, Argentina. This work is 
supported by ANPCyT PICT-2015-2752, Argentina. The author thanks the members of the Pierre Auger 
Collaboration, specially C. Dobrigkeit for reviewing the manuscript.

\end{document}